\documentclass[reprint,showpacs,
superscriptaddress,
 amsmath,amssymb,
 aps,prl
]{revtex4-1}

\usepackage{graphicx} 
\usepackage{dcolumn}
\usepackage{color}
\usepackage{bm}

\begin{document}

\title{Field correlations and the ultimate regime of turbulent convection}

\author{Mahendra K. Verma}
\email{mkv@iitk.ac.in}
\affiliation{Department of Physics, Indian Institute of Technology, Kanpur 208016, India}
\author{Pankaj K. Mishra}
\affiliation{Department of Physics, Indian Institute of Technology, Kanpur 208016, India}
\author{Ambrish Pandey}
\affiliation{Department of Physics, Indian Institute of Technology, Kanpur 208016, India}
\author{Supriyo Paul}
\affiliation{Computational Fluid Dynamics Team, Centre for Development of Advanced Computing, Pune 411 007, India}
\date{\today}

\begin{abstract}

Using direct numerical simulations of Rayleigh-B\'{e}nard convection (RBC)  under free-slip boundary condition, we show that the normalized correlation function between the vertical velocity field and the temperature field, as well as the normalized viscous dissipation rate, scales as $Ra^{-0.22}$ for moderately large Rayleigh number $Ra$.  This scaling accounts for the  Nusselt number ($Nu$) exponent  to be around 0.3 observed in experiments.    Numerical simulations also reveal that the above normalized correlation functions are constants for the convection simulation under periodic boundary conditions.
\end{abstract}

%\pacs{21.1.1} 

% PACS, the Physics and Astronomy
                             % Classification Scheme.
%\keywords{Suggested keywords}%Use showkeys class option if keyword
                              %display desired
% 47.27.te Turbulent convective heat transfer
% 47.55.P- Buoyancy-driven flows; convection
% 47.27.-i Turbulent flow
% 47.27.T- Turbulent transport processes
\maketitle

Thermal convection, which is ubiquitous in engineering and natural flows, exhibits various interesting phenomena, like instabilities, chaos, spatiotemporal patterns, and turbulence~\cite{Siggia:ARFM1994,Ahlers:RMP2009}.  Rayleigh-B\'{e}nard convection (RBC) is an idealized model of thermal convection in which a fluid is placed between two horizontal conducting plates, with the lower plate being hotter than the top one.  The dynamics of the flow is governed by two non-dimensional parameters: the Rayleigh number $Ra$, a measure of the strength of the buoyancy force, and the Prandtl number $Pr$, a ratio of the kinematic viscosity and the thermal diffusivity.   One of the most important theoretical and technological problems that remains unsolved in this field is nature of heat transport in RBC specially for very large Rayleigh number; this is the subject matter of this paper.

 Experiments and numerical simulations reveal certain universal properties for the heat flux~\cite{Siggia:ARFM1994,Ahlers:RMP2009,Kraichnan:PF1962b,Malkus:PRSL1954,Grossmann:JFM2000,Shraiman:PRA1990}.   Various experiments show that the Nusselt number, which is the ratio of the total (convective+conductive) heat flux to the conductive heat flux, scales as $Ra^{\beta}$ with $\beta$ around 0.3  for moderately large Rayleigh numbers~\cite{Castaing:JFM1989,Cioni:JFM1997,Niemela:NATURE2000,Glazier:NATURE1999,Urban:PRL2011,Ahlers:PRL2009,Chavanne:PF2001,Roche:PRE2001}.     However for very high Rayleigh number (called the ``ultimate regime"),  Kraichnan~\cite{Kraichnan:PF1962b} predicted that  $\beta = 1/2$; several experiments found no evidence of the ultimate regime~\cite{Castaing:JFM1989,Cioni:JFM1997,Niemela:NATURE2000,Glazier:NATURE1999,Urban:PRL2011,Ahlers:PRL2009}, while some others claimed its existence~\cite{Chavanne:PF2001,Roche:PRE2001}.  In this paper we show that the velocity-temperature correlation and the viscous dissipation rate vary with the Rayleigh number so as to yield the Nusselt number exponent to be 0.3 for the intermediate Rayleigh numbers.

One of the earliest attempt to understand Nusselt number scaling was by Kraichnan~\cite{Kraichnan:PF1962b} who derived  using the mixing-length theory that $Nu \propto Ra^{1/3}$ for large $Pr$, $Nu \propto (PrRa)^{1/3}$ for small $Pr$,  and  $Nu \sim 1$ for very small $Pr$.   For very large Rayleigh number or the ultimate regime, Kraichnan~\cite{Kraichnan:PF1962b} showed that $Nu \propto [ Ra/(\ln Ra) ]^{1/2}$.   Malkus~\cite{Malkus:PRSL1954} argued  that the $Nu$ exponent of 1/3 is due to the property that the heat flux is independent of the cell height. By separating the dissipation rates in the bulk and the boundary layers, Grossmann and  Lohse~\cite{Grossmann:JFM2000}  showed that for the bulk dominated convective flows,  $Nu \sim (PrRa)^{1/2}$ when $\lambda_u < \lambda_\theta$, but  $Nu \sim Ra^{1/3}$  when $\lambda_u > \lambda_\theta$.  Here $\lambda_u$ and $\lambda_\theta$ are the widths of the viscous and thermal boundary layers, respectively. The parameter space of validity for the above scaling has been detailed in Grossmann and  Lohse~\cite{Grossmann:JFM2000}.  Shraiman and Siggia~\cite{Shraiman:PRA1990}, Castaing {\em et al.}~\cite{Castaing:JFM1989}, and Cioni {\em et al.}~\cite{Cioni:JFM1997} computed the $Nu$ exponent and deduced it to be 2/7 due to the boundary layers.   Using scaling arguments Verzicco and Camussi~\cite{Verzicco:JFM1999} claimed that $Nu \sim (Ra)^{1/4}$ for small $Pr$ ($Pr < 1$).  
 
\begin{figure}[htbp]
 \begin{center}
 \includegraphics[scale=0.35]{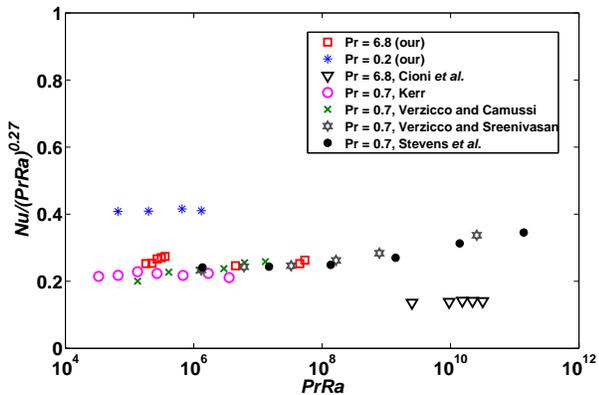}
 \end{center}
 \caption{Reduced Nusselt number ($Nu/(PrRa)^{0.27}$) versus $PrRa$, the product of Prandtl number ($Pr$) and Rayleigh number ($Ra$). Experimental data: Cioni {\it et al.}~\cite{Cioni:JFM1997} for water (black $\bigtriangledown$). Numerical data: Verzicco and Camussi~\cite{Verzicco:JFM1999} for $Pr = 0.7$ (green $\times$); Kerr~\cite{Kerr:JFM1996} for $Pr = 0.7$ (magenta $\bf{\circ}$); Stevens {\it et al.}~\cite{Stevens:JFM2010} for $Pr=0.7$ (black $\bullet$); Verzicco and Sreenivasan~\cite{Verzicco:JFM2008} for $Pr = 0.7$ (grey hexagons); our simulation data for $Pr=6.8$ (red $\Box$) and for $Pr=0.2$ (blue $\ast$).}  
 \label{fig:nu_pr}
 \end{figure}

Many experiments have been performed to test the above scaling, yet they have not been able to resolve the scaling exponents completely.   Laboratory experiments on typical fluids, water, helium gas, and mercury, yield Nusselt number exponents  from 0.26 to 0.31 for $Ra$ up to $10^{17}$~\cite{Castaing:JFM1989,Cioni:JFM1997,Niemela:NATURE2000,Glazier:NATURE1999,Urban:PRL2011,Ahlers:PRL2009}.     Cioni {\em et al.}~\cite{Cioni:JFM1997} and Glazier {\em et al.}~\cite{Glazier:NATURE1999} used mercury, while Castaing {\em et al.}~\cite{Castaing:JFM1989} and Niemela {\em et al.}~\cite{Niemela:NATURE2000}  used helium gas as their experimental fluid.  Cioni {\em et al.}~\cite{Cioni:JFM1997} also performed experiments on water for comparison.  Ahlers {\em et al.}~\cite{Ahlers:PRL2009} employed pressurized mixture of gas in their experiment.   Fig.~\ref{fig:nu_pr} illustrates the plots of reduced Nusselt number $Nu/(PrRa)^{0.27}$ vs.~$PrRa$ computed in earlier RBC experiments and numerical simulations.  Chavanne {\em et al.}~\cite{Chavanne:PF2001} and Roche {\em et al.}~\cite{Roche:PRE2001}  reported existence of the ultimate regime in their experiment on helium.  Roche {\em et al.} used nonsmooth surfaces to cancel the thickness variation of the viscous sublayer.    Ahlers {\em et al.}~\cite{Ahlers:PRL2009}  found simultaneous existence of multiple Nusselt number exponents (0.17, 0.25, and 0.36) for large $Ra$'s, possibly due to multiple coexisting attractors.
 
The Nusselt number for RBC is also investigated using direct numerical simulation.    Verzicco  and Camussi~\cite{Verzicco:JFM1999} and Kerr and Herring~\cite{Kerr:JFM2008}
showed that $Nu \sim Ra^{1/4}$ for small $Pr$, while Silano {\em et al.}~\cite{Silano:JFM2010}, Stevens {\em et al.}~\cite{Stevens:JFM2010}, and Verzicco and Sreenivasan~\cite{Verzicco:JFM2008} reported the exponent be around 1/3 for very high Rayleigh number.   Kerr~\cite{Kerr:JFM1996} and Kerr and Herring~\cite{Kerr:JFM2008} found the exponent to be around $0.28 \approx 2/7$ for larger $Pr$.   Fig.~\ref{fig:nu_pr} illustrates some of these results, as well as our numerical results for $Pr=0.2$ and 6.8 (to be described later). Kerr~\cite{Kerr:PRL2001} also studied the energy budget in RBC by computing the mean square velocity and dissipation rates as a function of Rayleigh number for various aspect ratios and Prandtl numbers.

For bulk-dominated convective flows, a simple argument would predict that  $Nu \propto \langle u_z \theta \rangle  \sim Ra^{1/2}$ ($\theta$= temperature fluctuation).   This scaling argument assumes that $\langle u_z^2 \rangle^{1/2} \sim Ra^{1/2}$ and $\langle \theta^2 \rangle^{1/2} \sim const$, and it ignores the correlation between the vertical velocity field and the temperature field.  Similarly, one of the exact relations $Nu-1 \sim  \epsilon^u /(PrRa)$~\cite{Shraiman:PRA1990} yields $Nu \sim Ra^{1/2}$ if we replace $\epsilon^u \sim U_L^3/L$ ignoring correlations  ($U_L$ is the large scale velocity, which is the free-fall velocity).  A major difficulty in this field is how to reconcile the $Nu \sim Ra^{0.3}$ scaling observed in the experiments to the $Nu \sim Ra^{1/2}$ predicted for the ultimate regime.   In this paper, we will explicitly compute the velocity-temperature correlation function and show that it scales with $Ra$ in such a way that $Nu \sim Ra^{0.3}$ at moderate Rayleigh numbers.   

Properties of convective flow depends quite critically on the boundary layers~\cite{Reeuwijk:PRE2008}.  
However, the $Nu$ scaling appears to be somewhat insensitive to the presence of boundary layers~\cite{Verzicco:EPL2003} and change of the boundary conditions~\cite{Verzicco:JFM2008}.   Motivated by the above observations, we attempt to compute the  scaling of the Nusselt and P\'{e}clet numbers in turbulent regime in terms of bulk quantities, specially the large scale velocity $U_L$ and the large scale temperature $\theta_L$ ($=\sqrt{\langle \theta^2 \rangle}$). The combined effects of the bulk and boundary layers would be studied later with more refined simulations and theoretical arguments.

The RBC equations under Boussinesq approximation for a fluid placed between two plates, separated by distance $d$ and with a temperature difference of $\Delta$, are
\begin{eqnarray}
\partial_{t}\mathbf{u} + (\mathbf{u} \cdot \nabla)\mathbf{u} & = & -\frac{\nabla \sigma}{{\rho}_0} + \alpha g \theta \hat{z} + \nu{\nabla}^2 \mathbf{u}, \label{eq:u} \\
\partial_{t}\theta + (\mathbf{u} \cdot \nabla)\theta & = & \frac{\Delta}{d} u_{z} + \kappa{\nabla}^{2}\theta, \label{eq:th} 
\end{eqnarray}
where $\theta$ and $\sigma$ are, respectively, the temperature and pressure fluctuations from the steady conduction state ($T = T_c + \theta$, with $T_c$ as the conduction temperature profile), $\hat{z}$ is the buoyancy direction, ${\rho}_0$ is the mean density of fluid, $g$ is the acceleration due to gravity, and $\alpha$, $\nu$, and $\kappa$ are the thermal heat expansion coefficient, the  kinematic viscosity, and the thermal diffusivity of fluid, respectively.   The two most important nondimensional parameters of RBC are the Rayleigh number $Ra = \alpha g \Delta d^3/\nu \kappa$ and  the Prandtl number $Pr = \nu/\kappa$.   The Nusselt number can be expressed as 
\begin{equation}
Nu = \frac{\kappa \Delta/d + \langle u_z T \rangle}{\kappa \Delta/d } = 1+\langle \frac{u_z d}{\kappa} \frac{\theta}{\Delta} \rangle = 1 + \langle u_z' \theta' \rangle,  \label{eq:Nusselt}
\end{equation}
where the nondimensionalized vertical velocity and temperature fields are $u_z' =u_z d/\kappa$  and $\theta' = \theta/\Delta$.  

Eq.~(\ref{eq:th}) has three competing terms: $(\mathbf{u} \cdot \nabla)\theta$, $\frac{\Delta}{d} u_{z}$, and $\kappa{\nabla}^{2}\theta$.  When the diffusion term $\kappa{\nabla}^{2}\theta$ of  Eq.~(\ref{eq:th}) is much smaller than the other two, we obtain 
\begin{equation}
\frac{U_L \theta_L}{d}  \approx  \frac{(u_z)_L \Delta}{d} \hspace{0.5cm} \Rightarrow \theta_L \approx\Delta.
\end{equation}
 In the momentum equation [Eq.~(\ref{eq:u})] 
\begin{equation}
\frac{U_L^2}{d} \approx \alpha g \theta_L  \hspace{0.5cm}  \Rightarrow  U_L \approx \sqrt{\alpha g \Delta d}, 
\end{equation}
implying that the P\'{e}clet number $Pe = U_L d/\kappa $ scales as
%\begin{equation}
$Pe   \approx \sqrt{PrRa}$.
%\end{equation}
These scaling relations are applicable when $(u_z)_L \Delta/d  \gg \kappa \nabla^2 \theta$ or $PrRa \gg 1$.   These results are consistent with the predictions of Kraichnan~\cite{Kraichnan:PF1962b},  Grossmann and  Lohse~\cite{Grossmann:JFM2000}, and Silano {\em et al.}~\cite{Silano:JFM2010}. 

\begin{figure}[htbp]
\begin{center}
\includegraphics[scale=0.35]{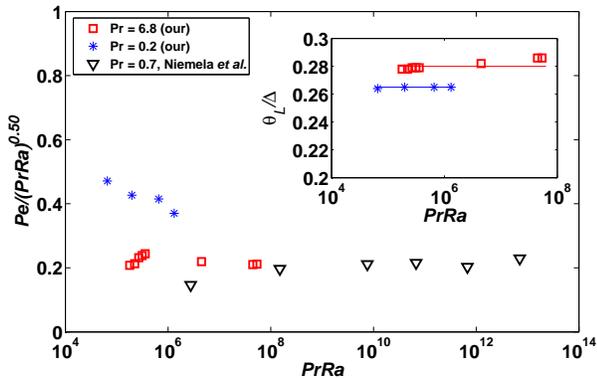}
\end{center}
\caption{ Reduced P\'{e}clet number ($Pe/(PrRa)^{0.5}$) versus $PrRa$. Black $\bigtriangledown$ represent the experimental data for helium ($Pr\sim0.7$) by Niemela {\it et al.}~\cite{Niemela:JFM2001}.  Red $\Box$ and blue $\ast$ represent our simulation data for $Pr=6.8$ and $Pr=0.2$, respectively. The inset shows the constancy of  the large scale temperature field ($\theta_L/\Delta$) with $PrRa$ for $Pr=6.8$  (red $\Box$) and $Pr=0.2$ (blue $\ast$).}
\label{fig:pe_pr}
\end{figure}

Note however that for $PrRa \ll 1$, the diffusion term dominates the nonlinear term, and it is balanced by $u_z \Delta/d$.  Therefore, $\theta_L  \approx U_L d \Delta/\kappa $ and $Pe \approx PrRa$.  Under this condition, $U_L \propto \theta_L$, hence $Nu = 1 + c (PrRa)^2$ with $c$ as a constant.   Convective turbulence with $PrRa \ll 1$  limit is very  rarely observed in terrestrial experiments or astrophysical observations.   Therefore, in the present paper we limit ourselves to $PrRa \gg 1$ regime.    In Fig.~\ref{fig:pe_pr} we plot the reduced P\'{e}clet number $Pe/(PrRa)^{0.5}$ computed in various RBC experiments and numerical simulations; these results are in general agreement with $Pe \sim (PrRa)^{1/2}$ scaling~\cite{Niemela:JFM2001}. The data of Figs.~\ref{fig:nu_pr} and \ref{fig:pe_pr}  appear to be compactified reasonably well with $PrRa$, hence we use $PrRa$ as a scaling variable.   

As discussed earlier, a naive replacement of $\langle u_z' \theta' \rangle$ of Eq.~(\ref{eq:Nusselt}) with $\langle u_z'^2 \rangle^{1/2}  \langle \theta'^2 \rangle^{1/2}$ yields
$Nu \sim \sqrt{PrRa}$, which is not observed in the experiments and simulations for moderately large Rayleigh numbers (see Fig.~\ref{fig:nu_pr}).  To account for the velocity-temperature correlation, we rewrite the Nusselt number as
\begin{equation}
Nu - 1  = \langle  u_z' \theta' \rangle =  C^{u\theta}(PrRa)  \langle  u_z'^2  \rangle^{1/2}_V \langle \theta'^2 \rangle_V^{1/2},
 \label{eq:Nu2}
 \end{equation}
where  $C^{u\theta}(PrRa) =  \langle \frac{\langle u_z' \theta' \rangle_V} {  \langle u_z'^2  \rangle_V^{1/2}   \langle  \theta'^2  \rangle_V^{1/2}}  \rangle_t$ is the normalized correlation function between the vertical velocity and temperature, with  $\langle \rangle_V$ and  $\langle \rangle_t$ representing the volume and the temporal averages, respectively.  Note that $C^{u\theta}(PrRa) \le 1$ as a consequence of Cauchy-Schwarz inequality.   We perform numerical simulations to compute the above correlation, along with the P\'{e}clet number,  Nusselt number, and $\theta_L$,  for various $Pr$ and $Ra$ values.  

We numerically solve Eqs.~(\ref{eq:u}, \ref{eq:th}) using pseudo-spectral method on grid-size ranging from $128^3$ to $512^3$.  We apply free-slip boundary conditions for the velocity field and isothermal boundary condition for the temperature field.  We use Runge-Kutta fourth-order scheme for time advancement, and compute the relevant quantities in the steady state. For further details of the simulation, refer to~\cite{Mishra:PRE2010,Verma:Arxiv2011}.  

In Figs.~\ref{fig:nu_pr} and~\ref{fig:pe_pr} we plot the computed reduced Nusselt  and P\'{e}clet numbers vs.~$PrRa$  for $Pr=6.8$ (red squares) and 0.2 (blue circles).  The best fits for our data are $Nu = (0.27\pm0.04)(PrRa)^{0.27\pm0.01}$ and $Pe = (0.26\pm0.04)(PrRa)^{0.49\pm0.01}$ for $Pr=6.8$, and  $Nu = (0.39\pm0.02)(PrRa)^{0.27\pm0.01}$ and $Pe= (1.04\pm0.20) (PrRa)^{0.43\pm0.02}$ for $Pr=0.02$.   The inset of Fig.~\ref{fig:pe_pr} demonstrates constancy of $\theta_L/\Delta$ with $PrRa$.  These results are consistent with the earlier experimental and numerical observations, as shown in the figures.

\begin{figure}[htbp]
\begin{center}
\includegraphics[scale=0.30]{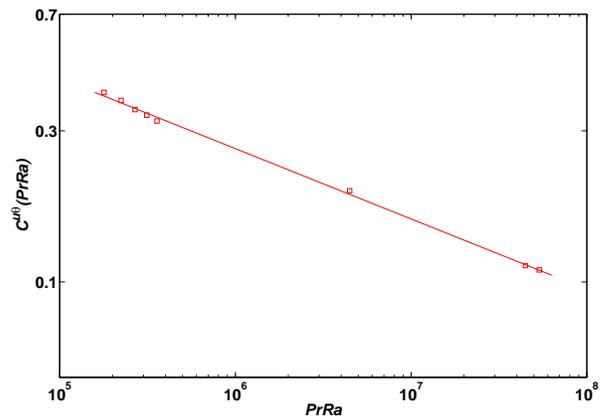}
\end{center}
\caption{Normalized correlation function $C^{u\theta}(PrRa) =  \langle \langle u_z' \theta' \rangle_V / (  \langle u_z'^2  \rangle_V^{1/2}   \langle  \theta'^2  \rangle_V^{1/2})  \rangle_t$ versus $PrRa$. Function $(5.6 \pm 1.30)  (PrRa)^{-0.22 \pm 0.017}$ fits well with the simulation data.}
\label{fig:cos_zeta}
\end{figure}

We  compute the  normalized correlation function $C^{u\theta}(PrRa)$ for $Pr=6.8$ and various $Ra$ values, and plot them in Fig.~\ref{fig:cos_zeta}. We observe that
\begin{equation}
C^{u\theta}(PrRa)  \approx (5.6 \pm 1.30)  (PrRa)^{-0.22 \pm 0.017}
\label{eq:cos_zeta}
\end{equation}
for $Ra \lessapprox 10^8$.  The above correlation is due to the interactions between the bulk flow and the boundary layers.  However its power law behaviour is interesting which immediately yields $Nu \sim Ra^{0.27}$, consistent with the $Nu$ scaling observed in experiments and numerical simulations.  Also, a combination of Eqs.~(\ref{eq:Nu2}) and (\ref{eq:cos_zeta}) provides
\begin{equation}
Nu-  1 =  0.046 \{ C^{u\theta}(PrRa) \}  (PrRa)^{1/2}.
\label{eq:Nu3}
\end{equation}

We can also deduce another important field correlation using one of the exact relations for the RBC flow~\cite{Shraiman:PRA1990}:
\begin{equation}
Nu-1  =   \frac{Pr^2 d^4 \epsilon^u}{\nu^3 Ra} = \frac{(Pe)^3}{PrRa} C^{\epsilon^u}(PrRa),
\label{eq:Nu_eps1}
\end{equation}
where the dissipation rate of the kinetic energy $\epsilon^u = (U_L^3/d) C^{\epsilon^u}(PrRa)$, with the function $C^{\epsilon^u}(PrRa)$, named as ``normalized viscous dissipate rate", playing similar role as $C^{u\theta}(PrRa)$.  Since $Pe = a (PrRa)^{1/2}$, $C^{\epsilon^u}(PrRa)$ must scale as $s (PrRa)^{-0.22}$ similar to $C^{u\theta}(PrRa)$ (Here $a,s$ are constants).  Thus
\begin{equation}
Nu-1  =   a^3 (PrRa)^{1/2} s(PrRa)^{-0.22} = a^3 s (PrRa)^{0.27}
\label{eq:Nu_eps2}
\end{equation}
that immediately yields $s \approx 15$.  

\begin{figure}[htbp]
\begin{center}
\includegraphics[scale=0.30]{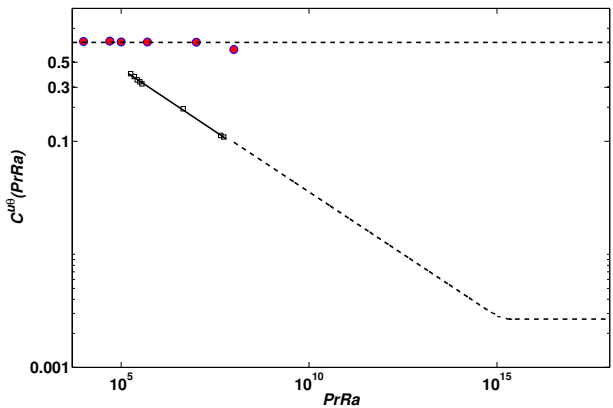}
\end{center}
\caption{The purple circles with extended chained line depict constancy of the normalized correlation function $C^{u\theta}(PrRa)$ for the periodic box  ($Pr=1$). Assuming that the convective turbulence becomes fully-developed for very large $Ra$, we conjecture that the normalized correlation function would become a constant in the ultimate regime.  The second curve is reproduction of the correlation function of Fig. 3 along with extended $Ra^{-0.22}$ for moderately large $Ra$, and then a constant for the ultimate regime after some transitional Rayleigh number (here $10^{15}$). 
 }

% {\color{red}{Assuming that the convective turbulence becomes fully-developed for very large $Ra$, we conjecture that the normalized correlation function would become a constant in the ultimate regime.  The second curve is reproduction of the correlation function of Fig. 3 along with extended $Ra^{-0.22}$ for moderately large $Ra$, and then a constant for the ultimate regime after some transitional Rayleigh number (here $10^{12}$~\cite{Roche:PRE2001}). }}
\label{fig:cos_zeta_periodic}
\end{figure}

The normalized  correlation  function $C^{u\theta}(PrRa)$ and the normalized viscous dissipation rate $C^{\epsilon^u}(PrRa)$  decrease with increasing $Ra$  since  the fields tend to become more and more turbulent. A question is whether these correlations would continue to decrease and they will vanish as $Ra \rightarrow \infty$, or  they would saturate to some asymptotic values.   For very large Rayleigh numbers, we expect the turbulence to be fully developed, and, according to the Kolmogorov's theory for fully-developed turbulence, $\epsilon^u \sim (U_L^3/L)$ or $C^{\epsilon^u}(PrRa) \sim const$.  This would correspond to the ``ultimate regime" proposed  by Kraichnan~\cite{Kraichnan:PF1962b}.   Therefore, we conjecture that the functions $C^{\epsilon^u}(PrRa)$ or $C^{u\theta}(PrRa)$ would be as shown in Fig.~\ref{fig:cos_zeta_periodic}; a power law for intermediate $Ra$, and constant for very large $Ra$ (ultimate regime).  

The constant values of the $C^{u\theta}(PrRa)$  and $C^{\epsilon^u}(PrRa)$ in the predicted ultimate regime can be computed given the transitional Rayleigh number beyond which the ultimate regime is expected.   From the present set of experiments~\cite{Castaing:JFM1989,Cioni:JFM1997,Verzicco:JFM1999,Niemela:NATURE2000,Glazier:NATURE1999,Urban:PRL2011,Ahlers:PRL2009,Chavanne:PF2001,Roche:PRE2001}, we can conclude that the ultimate regime possibly starts beyond $Ra \approx10^{17}$\cite{Roche:PRE2001}.  With this Rayleigh number, the asymptotic values of $C^{u\theta}(PrRa)$  and $C^{\epsilon^u}(PrRa)$ would be around $0.00066$ and $0.0018$, respectively.  Using these values, we conjecture that the Nusselt number in the ultimate regime would scale as $Nu \approx 3.1\times 10^{-5} (PrRa)^{1/2}$.

 It has been conjectured that turbulent convection of the ultimate regime is approximately represented by  convection with {\em periodic boundary conditions} at significantly lower Rayleigh numbers~\cite{Lohse:PRL2003}.  To strengthen our theoretical arguments presented earlier, we carried out RBC simulations for the periodic box geometry of size $(2\pi)^3$ using a pseudospectral code~\cite{Mishra:PRE2010,Verma:Arxiv2011}.  We computed the normalized correlation function $C^{u\theta}(PrRa)$  for $Pr=1$ and high Rayleigh numbers $Ra=10^4$--$10^8$, for which the flow is fully turbulent.  We observe that  for all the runs $C^{u\theta}(PrRa)  \approx 0.75$ as shown in Fig.~\ref{fig:cos_zeta_periodic}, as well as $C^{\epsilon^u}(PrRa) \approx 0.48$. Also, $Pe \approx 5.7 Ra^{0.50 \pm, 0.02}$, $Nu \approx 23.7 Ra^{0.46\pm0.04}$, and $\theta/\Delta \approx 4.21\pm 0.21$.   Moreover, the flow is highly anisotropic with $\langle 2 |u_z|^2/( |u_x|^2 + |u_y|^2) \rangle \approx 3.26\pm0.43$.  Thus, our prediction for the ultimate regime is consistent with the convective flow with periodic boundary conditions.

In summary, we relate the Nusselt number scaling to the normalized  velocity-temperature correlation function, as well as the normalized viscous dissipation rate.  We show that these functions scale with the Rayleigh number as $Ra^{-0.22}$ for intermediate $Ra$'s, which leads to  $Nu \sim Ra^{0.3}$ observed in experiments.   For very large Rayleigh numbers, the flow is expected to be fully turbulent.  This observation leads us to conjecture that the normalized velocity-temperature correlation function could become a constant for very large $Ra$, thus yielding $Nu \sim Ra^{1/2}$, as predicted by Kraichnan~\cite{Kraichnan:PF1962b}.   Convection simulations for a periodic box geometry also exhibit constant values for the normalized correlation functions, as well as $Nu \sim Ra^{1/2}$, consistent with our conjecture.  Thus, the field correlations  provide valuable insights into the scaling of the Nusselt number.  

Numerical simulation of ultimate regime is beyond the capabilities of present computers, but   gap is expected to be bridged soon, which would settle this long-standing problem.  Also, attempts to compute the asymptotic values of the correlations functions in the similar lines as the theoretical computations of Whitehead and Doering, and Constantin and Doering~\cite{Whitehead:PRL2011} would be very valuable.

\begin{acknowledgments}
We thank Stephan Fauve and Mani Chandra for very useful discussions, and Computational Research Laboratory, Pune, and Centre for Development of Advanced Computing, Pune for the  computing time on EKA and PARAM YUVA, respectively.  Part of the computation was also done at HPC cluster of IIT Kanpur.  Part of this work was supported by Swarnajayanti fellowship to MKV, and a BRNS grant.
\end{acknowledgments}

%\bibliographystyle{apsrev}
%\bibliography{/Users/mkv/res/bib/turbulence.bib}

\end{document}